# An Internet of Things Framework for Smart Energy in Buildings: Designs, Prototype, and Experiments

Jianli Pan[*], Raj Jain[‡], Subharthi Paul[◊], Tam Vu[§], Abusayeed Saifullah[#], Mo Sha[※]

[*]University of Missouri – St. Louis, pan@umsl.edu
[‡]Washington University in St. Louis, [◊] Cisco Systems, Inc.
[§]University of Colorado – Denver, [#]Missouri University of Sci. & Tech.
[※]Division of Science, Governors State University

***Abstract*— Smart energy in buildings is an important research area of Internet of Things (IoT). Buildings as important parts of the smart grids, their energy efficiency is vital for the environment and global sustainability. Using a LEED-gold-certificated green office building, we built a unique IoT experimental testbed for our energy efficiency and building intelligence research. We first monitor and collect one-year-long building energy usage data and then systematically evaluate and analyze them. The results show that due to the centralized and static building controls, the actual running of green buildings may not be energy efficient even though they may be "green" by design. Inspired by "energy proportional computing" in modern computers, we propose a IoT framework with smart location-based automated and networked energy control, which uses smartphone platform and cloud computing technologies to enable multi-scale energy proportionality including building-, user-, and organizational-level energy proportionality. We further build a proof-of-concept IoT network and control system prototype and carried out real-world experiments which demonstrate the effectiveness of the proposed solution. We envision that the broad application of the proposed solution has not only led to significant economic benefits in term of energy saving, improving home/office network intelligence, but also bought in a huge social implication in terms of global sustainability.**

## I. Introduction

Smart energy in buildings is an important research area of Internet of Things (IoT). Buildings as important parts of the smart grids, their energy efficiency is vital for the environment and global sustainability. According to a general survey [1], in United States, buildings are responsible for around 38% of the total carbon dioxide emissions; 71% of the total electrical energy consumption; 39% of the total energy usage; 12% of water consumption; 40% of non-industrial waste. In the meantime, cost of traditional fossil fuels is rising and its negative impacts on the planet's climate and ecological balance make it important for us to explore new clean-energy sources and improve the energy efficiency in the consumer-side smart grids of various buildings.

However, buildings are complex systems and many factors can affect the total energy consumption in different buildings. Also, conventional buildings are not with too many intelligent designs. It is meaningful to monitor the real energy consumption data and find the major factors and patterns through systematic modeling and analysis for different types of buildings. Such results can be used to further design and implement appropriate IoT based networking system to construct appropriate methods and strategies improving the energy efficiency for both "green" and "non-green" (conventional) buildings. We summarize the research on the topic into three sequential key aspects:

(1) **Energy Monitoring**: Through communication networks, the consumption and generation of energy are monitored and logged in different granularities including the whole building, floors, departments, labs, rooms, and even occupants.

(2) **Energy Modeling and Evaluation**: Through off-line modeling and evaluation, identify the energy consumption patterns and factors that may influence the consumption and the extent of their impact.

(3) **IoT System to Apply Practical Changes and Strategy Adjustments**: The modeling and evaluation results are used to identify the key energy components of the building, to apply adjustments, and to devise strategies to reduce energy consumption. IoT based networking system is designed and prototyped to realize the strategies and achieve the goal.

Our research covers all the three aspects. We monitored and collected the building energy usage data for almost a year. Based on our data traces, we systematically identified the energy consumption patterns and explored potential methods to improve the energy efficiency. The results show that due to centralized and fixed pattern control, the actual running of green buildings may not be energy efficient even though they may be "green" by design. Inspired by "energy proportional computing" in modern computers, we propose a smart location-based automated energy control IoT framework using smartphone platform and cloud computing technologies to enable smart mobile control and multi-scale energy proportionality, which includes ***building-, user-, and organizational-level energy proportionality***. We further built an experimental IoT prototype system to demonstrate the effectiveness of our proposed idea. Our results show potential economic and social sustainability benefits.

Unlike simulation based solution, our work is based on real measured data traces for a currently in-use on-campus green building, and a real IoT system to control the energy automation. We use the latest information technologies such as mobile smartphones with location service, distributed control, and cloud computing to actively involve the occupants in the energy-saving process. Energy-saving policies from multiple sources such as individuals and organizations are considered in an integrated policy framework in deciding the final energy saving strategies. We aim to create an energy-efficiency IoT testbed that can be easily migrated to all kinds of buildings and achieve energy savings in multiple scales.

In this journal version paper, we first summarize and refine our previous work in two conference papers [2, 3]. In [2], we evaluated the building energy usage data and presented our findings in identifying the major issues in these buildings. Based on that, we proposed a smart location-based networked energy control IoT system design to tackle the issue and improve the energy efficiency [3].

In this paper, however, we add new contributions to complete the three steps described above. Particularly,

(1) We synthesize the previous separate contributions into a complete IoT framework design. It includes research and work in the whole process of identifying the key problems, finding methods to solve them, and developing prototype system to prove the effectiveness of the proposed method.

(2) We build a novel experimental prototype IoT system which demonstrates the real time location-based automated energy policy control across multiple buildings. It is the basic step in changing from the current centralized control and static energy consumption modes to distributed and dynamic energy control in the consumer-side smart grids containing various common buildings.

(3) Based on these, we propose to create a future of multi-scale energy proportionality. The central idea is to generalize the smartphone and location-based energy control idea and include policies of multiple levels of organizations. It aggregates the energy saving of individual users and allows distributed and dynamic energy control, which is the key for energy proportionality.

The rest of this paper is organized as follows. Section II presents the testbed description, methodology, and detailed energy efficiency data analysis and discussions. Section III describes our idea on the smart location-based automated energy control IoT architecture. Section IV discusses the prototype system and the experimental results, and the discussions on multi-scale energy proportionality. Section V reviews some related work and Section VI concludes the paper.

## II. ENERGY EFFICIENCY EVALUATION AND ANALYSIS

In this section, we describe our energy monitoring testbed, evaluation methodology, and data modeling and analysis results. This section is mostly based on our contributions in [3].

### *A. Energy Monitoring Testbed and Justification*

We realize that buildings can be very different from each other and it is extremely important to find the common "thing" or pattern among them in terms of energy efficiency. So in our project we talked to some on-campus building maintenance experts thoroughly and investigated the common structure of these buildings in U.S. For our testbed, we pick a very typical office building constructed in 2010 (actually two latest on-campus buildings constructed in 2014 also show that they share exactly the same technologies as our testbed due to the reason that traditional building-based systems are not an area that is developing as fast as those IT technologies). Besides our experimentation on this particular office building, we also investigated a case in another extreme end, a Net-Zero Energy Building named "Tyson Research Center", which is a small office building. It received prestigious award in the "living building challenge". We found that for such small office buildings or home buildings, it is relatively simpler and easier to apply networking technologies to control or change their energy policy. In comparison, large buildings like our testbed are more difficult to change and it is also one of the reasons why in this paper we primarily focus on such large office buildings. With our findings in this testbed, it is relatively easier to tailor and generalize the system to solve the issue with other buildings of the same type or different types.

Our testbed building received a Gold certificate from LEED rating system [4] by U.S. Green Building Council (USGBC) [5]. It adopts a series of energy efficiency and sustainability features. The overall resource usage data for the building are monitored and logged through a series of meters every 30 minutes (some are 15 minutes) through wired network for future off-line data modeling and analysis. It is a very ***typical*** large green office building with typical monitored subsystems such as HVAC (Heating, Ventilation, Air-conditioning, and Cooling), lighting, and water systems. We believe that the experiments and further data analysis findings from this testbed apply to other large office buildings.

### *B. Data Source and Analysis Methodology*

We sorted out the most useful measured data by analyzing the relationships among various parameters. Based on it, the data points that we use include: *the total electrical energy consumption, the heating and cooling energy consumption, and the outdoor and indoor environmental data such as temperature and humidity*. The heating and cooling parts are deemed as the HVAC consumption while the total electricity consumption covers a wider range of loads in the building. Though separate lighting data may be useful, such data is not currently available. Moreover, we unify the semi-hourly or hourly logged data to an hourly basis for uniform analysis.

Our primary modeling and evaluation goal is to identify the energy consumption pattern and know how it is related to: (1) ***environmental factors,*** and (2) ***occupancy rate.*** So, we first analyze the relationship between electricity, heating, and cooling energy consumption and the outdoor environmental factors. Our method is to combine the short period (longer than 1 day and less than 1 week) and the long period (several months) correlation analysis to show the overall trends. We group the hourly data into multiple granularities such as weekly and monthly to reveal the complete correlation differences over a relatively long period. We also develop Multiple Polynomial Regression (MPR) model and Multiple Linear Regression (MLR) model to reveal longer term average seasonality trends.

### *C. Detailed Evaluation and Analysis*

In this section, we present the detailed modeling and evaluation results and the corresponding analysis.

#### *1) Environmental Impacts Analysis*

Here, we focus on temperature and humidity, and study their impacts on the total electrical and HVAC energy consumption.

*a) Short Period Basic Trend and Correlation Analysis*

We put two groups of factors together: (1) group 1 made of electric consumption, heating energy, and cooling energy; (2) group 2 containing temperature and humidity. We want to see if there is any straightforward connection. Fig. 1 shows the relationship between electrical energy consumption and temperature. It shows almost no correlation.

Fig. 2 shows the relationship between heating energy and temperature, in which we still do not find very strong correlation (due to the similar patterns between heating and cooling, in this paper, we only show results of heating data). Note that in the figures we use the British Thermal Unit (BTU) as the unit for heating and cooling. 1 BTU is equal to 1055 joule or 0.293 watt-hours.

**Observation**: (1) Overall, correlations between both heating and electric energy consumption and the outdoor weather conditions are small. (2) Overall, the electrical consumption also shows very little variation between days and nights, which means that ***it possibly has a small correlation with occupancy.***

*b) Long Period Correlation Analysis*

We now study the correlation among multiple factors over a longer period. After filtering out incomplete and inaccurate data, we get a continuous dataset for about 10 months (39 weeks). It ranges from 3/18/2011 to 12/31/2011. We group the data into weeks and every week has 24*7=168 data points. For each 168 data point set, we calculate the correlations among multiple factors. These factors include: temperature (denoted as X), humidity (Y), total electrical energy consumption (Z), heating energy (H), and cooling energy (C). We also mark the seasons according to the Missouri climate convention.

The correlations between electrical energy consumption and weather conditions are shown in Fig. 3. They are mostly below 0.5. Interestingly, the correlation for summer season is a bit higher than that for fall and winter seasons. The results validate the results we showed in Fig.1 and 2. Note that the X and Y in the figures do not mean x-axis and y-axis, but temperature and humidity in our notation. The results for the correlations of heating energy with weather conditions are shown in Fig. 4.

**Observation**: The figures roughly indicate that ***the heating and cooling systems do not actively take the outdoor weather condition as factors to dynamically adjust the running schedule and policies to save energy***.

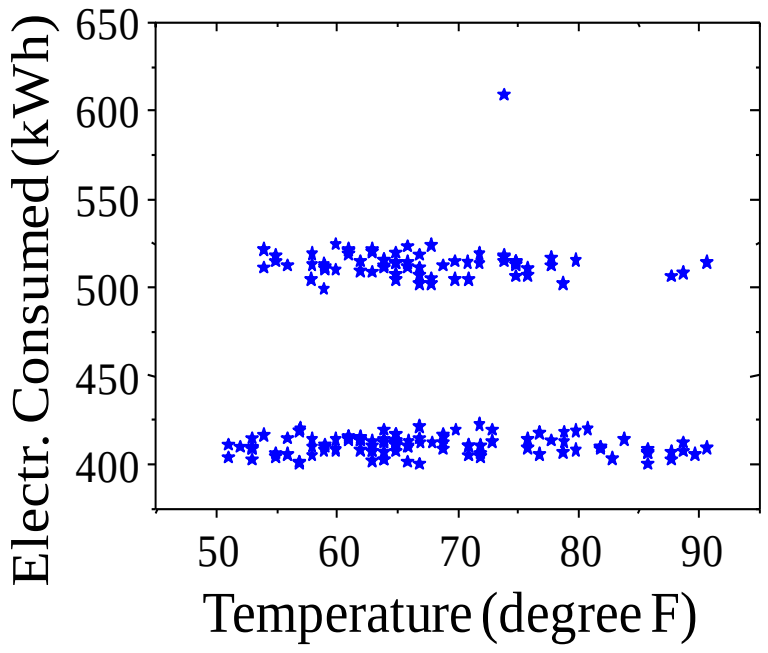

Fig. 1. Total **electrical** energy consumption with temperature

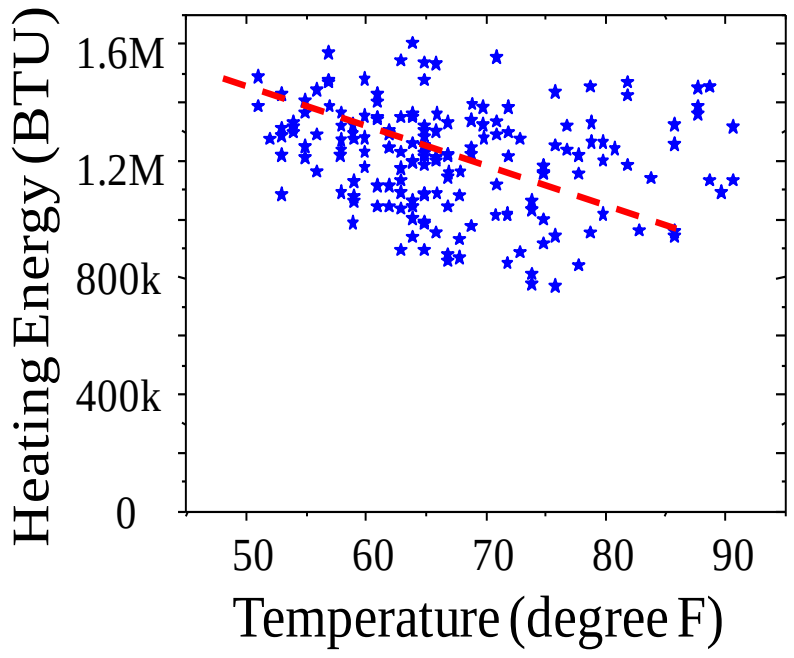

Fig. 2. **Heating** energy consumption with temperature

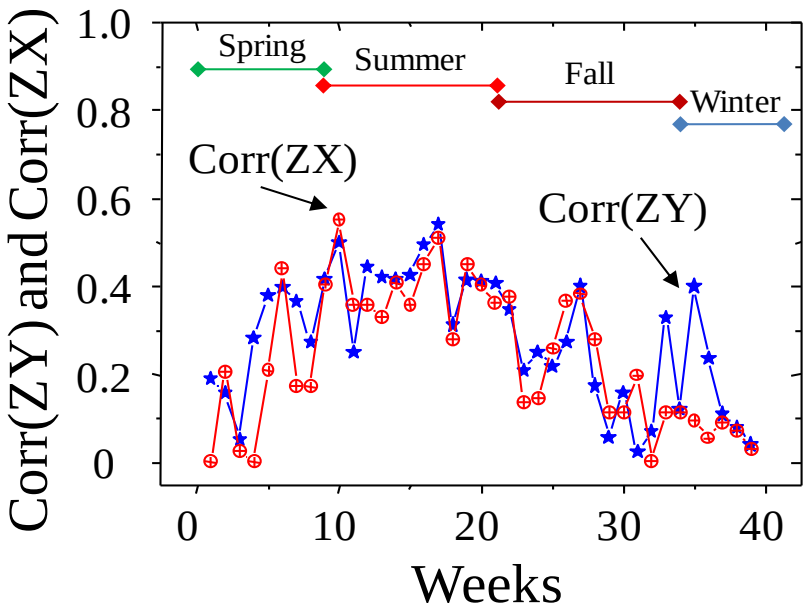

Fig.3. Correlations between electrical energy (Z), temperature (X), and humidity (Y)

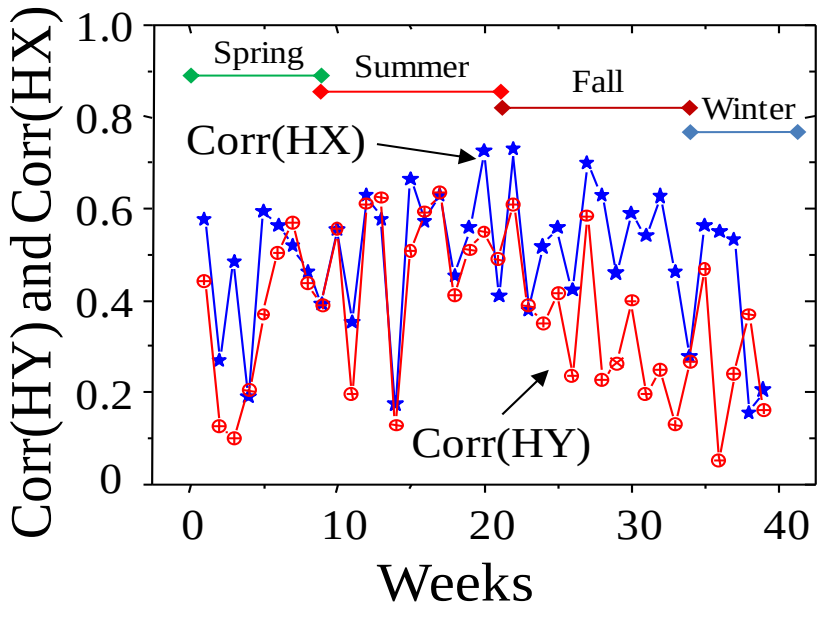

Fig. 4. Correlations between heating energy (H), temperature (X) and humidity (Y)

*c) Daily Average Data Analysis*

So far, we studied ***hourly*** electricity, heating, and cooling energy data (1 data sample per hour). We also aggregate the data into ***daily*** averages to see if there are any new findings. Specifically, we calculate the daily average temperature and humidity, and the daily total electric, heating, and cooling energy consumption. Then we have a data set for each day and a total 245 data sets from 5/1/2011 to 12/31/2011.

The daily heating and cooling trends are shown in Fig. 5. The seasonality is clear for both heating and cooling data in that there is more cooling and less heating energy in the summer. In total, for the above period, the energy usage is 8.1 billion BTU for heating and 16.9 billion BTU for cooling. It is interesting that the cooling system uses about ***twice*** the energy than heating. In the ***summer months the cooling energy usage is significantly higher than that in other months***. The daily electrical energy consumption in Fig. 6 shows a very regular fluctuation. Seasonality is not that obvious.

**Observation**: the electricity provisioning in this building is relatively fixed and “extra capacity” is generally provided to satisfy any burst usage. ***In other words, a lot of electrical energy is wasted, especially, during afterhours***.

*d) Regression Modeling and Analysis*

We further use regression models to analyze the relationship among multiple factors and use the statistical approaches to examine whether they can justify our findings. We try both Multiple Polynomial Regression (MPR) and Multiple Linear Regression (MLR) models, and compare the two results.

First, we use the same daily average dataset and we have a vector of data point for each day. The vector is <daily average temperature, daily average humidity, daily electrical energy, daily heating energy, daily cooling energy> and we have 245 data vectors in total. We compute the coefficients of each factor in the two types of regression models, calculate the errors and conduct tests to check the models' effectiveness.

Table I presents results of the MPR and MLR on electrical, heating, and cooling energy predictions with temperature and humidity as two parameters. As shown in Table I, the coefficient of determination $R^2$ is the fraction of the total variation explained by the regression [6]. For example, for electrical energy MPR, ***$R^2$ is 0.1902 which means that the MPR regression model can only explain 19.02% of the variation of electrical energy usage***. In comparison, the $R^2$ value for cooling energy is 0.9884 which means that the ***MPR can explain 98.84% of the variation of the cooling energy consumption***. This result validates our previous conclusion.

**Observation**: The regression reminds us that various energy subsystems of the buildings are impacted differently by the environmental factors; hence, for better energy efficiency, we should ***tune each subsystem separately***. For example, heating and cooling respond more to the environment and we may use environment condition to tune the running policy of the HVAC system and save energy.

*2) Occupancy Impact Analysis*

In this section, we focus on the occupancy and study how it can impact the energy consumption.

*a) Weekdays/Weekends Energy Comparisons*

We roughly divide the data into three subsets: *regular office hours* (8:00am to 8:00pm of weekdays), *after hours* (8:00pm to 8:00am of weekdays), and *weekend* (whole days of Saturday and Sunday). We study the data by weeks and for every week we have three subsets. For each subset, we calculate their electrical and heating energy averaged in 24 hours, and compare them to see the differences. The results are shown in Fig. 7 and Fig. 8, respectively. From Fig. 7, we can see that the electrical energy consumption during office hours is about 15% more than that for afterhours and weekends. The numbers for after hours and weekends ***are not as low as expected*** which also illustrates that the current building operation is far from efficient and is not proportional to the actual usage or occupancy.

The heating energy pattern shown in Fig. 8, however, is a little bit different. Overall, the heating energy consumption for afterhours is about 6% higher than weekends, and 19% higher than those for business hours.

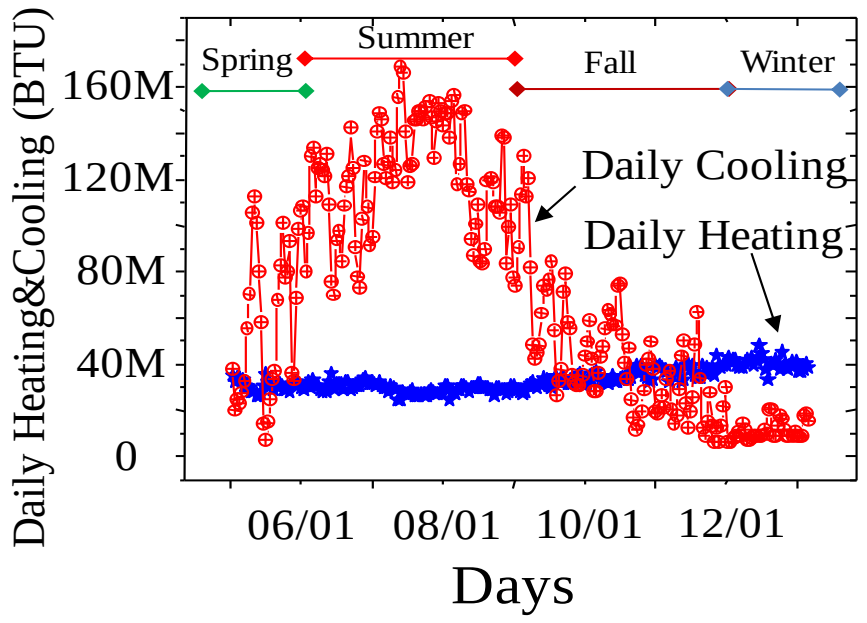

Fig. 5. Daily heating and cooling

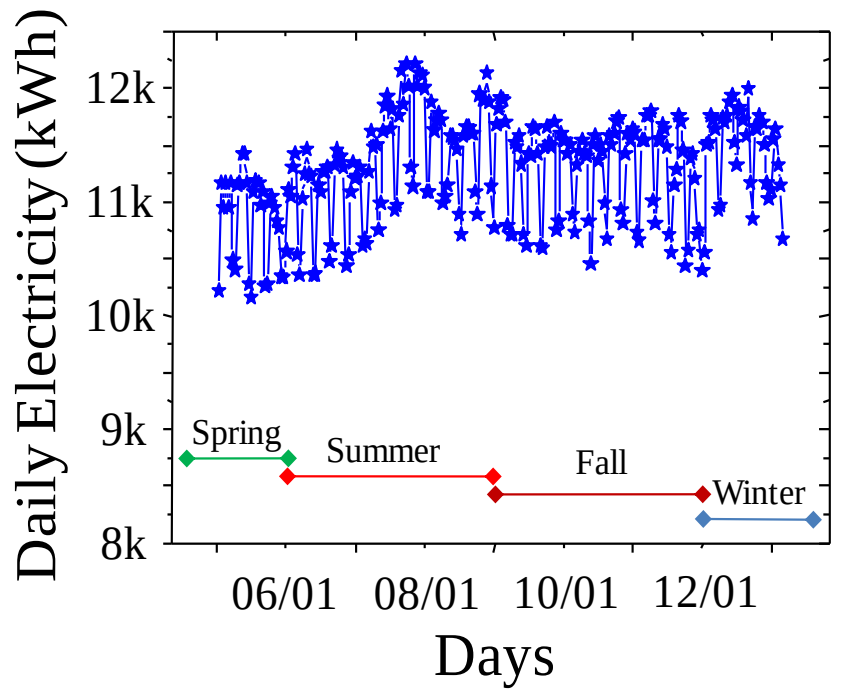

Fig. 6. Daily electrical energy consumption

Table I. Regression results for electricity, heating, and cooling energy

| | Electrical Energy | | Heating Energy | | Cooling Energy | |
|---|---|---|---|---|---|---|
| MPR | **$R^2$** | **0.1902** | **$R^2$** | **0.8634** | **$R^2$** | **0.9884** |
| MLR | **$R^2$** | **0.0213** | **$R^2$** | **0.8610** | **$R^2$** | **0.9072** |

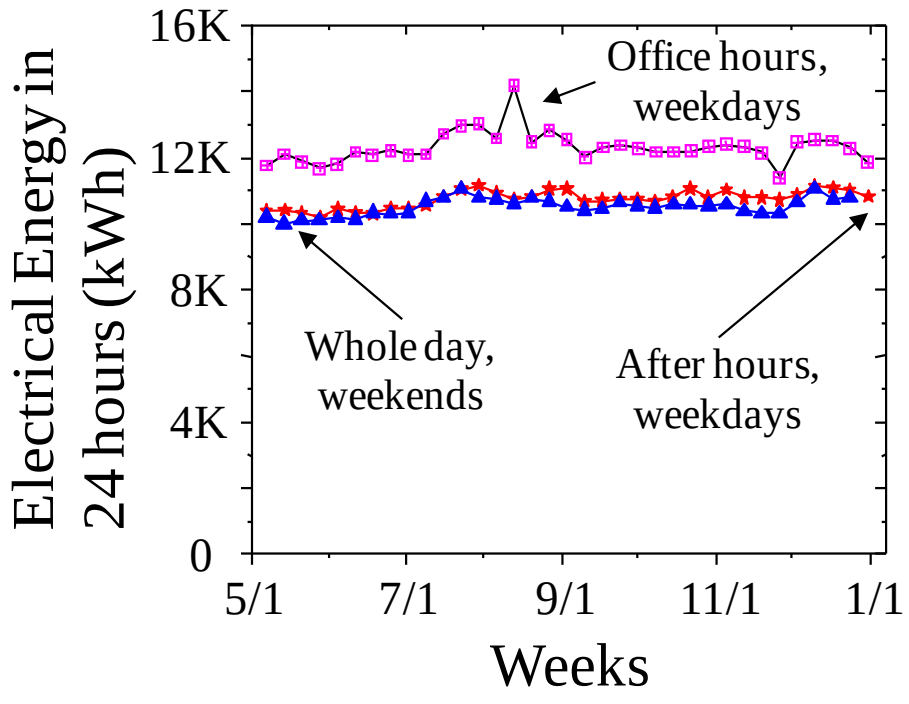

Fig. 7. Comparison of **electrical** energy consumption averaged in 24 hours for office hours, after hours, and weekends

It is interesting to know that the heating consumption for the business hours is the lowest compared to the other two. It is probably due to the fact that ***the occupancy rate is higher during the office hours and more people are active and providing body heat in the building and hence reduce external heating energy demands***.

**Observation**: the analysis clearly shows that the ***actual occupancy rate has very low impact to the energy***

***consumption***. Ideally, the numbers for after hours and weekends should be much lower than those for office hours.

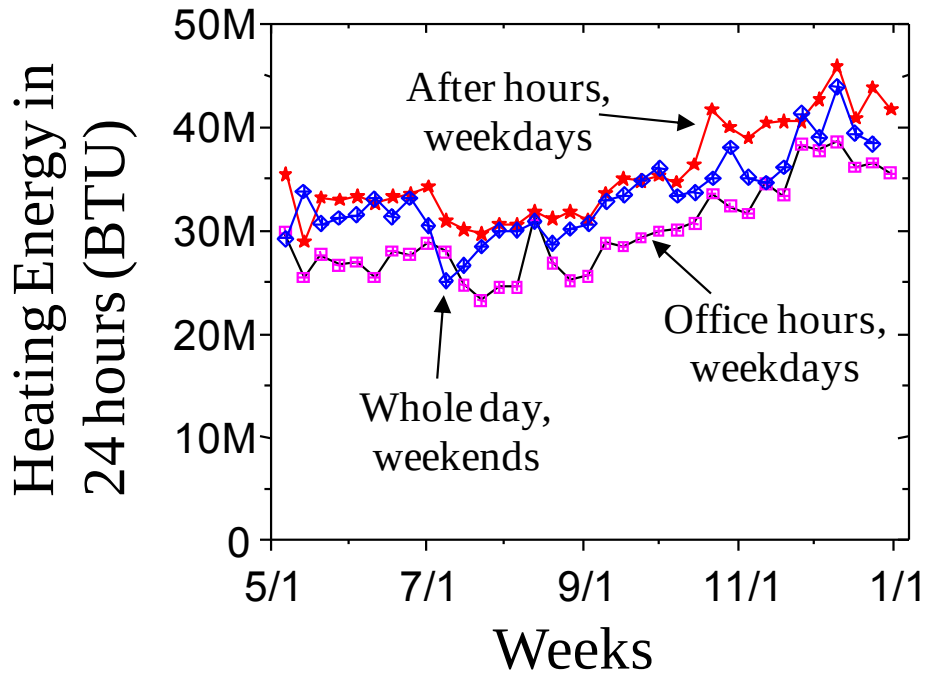

Fig. 8. Comparison of **heating** energy consumption averaged in 24 hours for office hours, after hours, and weekends

*b) In-semester/Holidays Energy Comparison*

To further see the occupancy rates' impact, we selected out the data for in-semester days and summer holidays. Based on the academic calendar of Washington University for year 2011, we pick the period between Aug. 30 to Dec. 09 (101 days in total) as the fall semester, and the period between May 10 to August 29 (111 days in total) as the summer holiday season. Generally, the summer holiday season has lower occupancy rate than the regular fall semester in our testbed.

Firstly, we compute the total electrical energy consumption for the above two periods and average them by the days. The results are shown in Fig. 9 which indicates that the electrical consumption varies very little for these two periods. We also compared the heating and cooling energy consumptions which are shown in Fig. 10. Daily average heating energy of the fall season is about 20% higher than summer, while daily cooling is 65% lower than summer season. Such results are consistent with the analysis results of the previous several subsections. If we separate the two periods into office hours and after hours, then we have a more detailed view of the energy consumption patterns. As shown in Fig. 11, we scale the daily energy consumption in Y axis into a 0 to 100 range. We find that for electricity usage during after hours, it is almost fifty-fifty between summer and fall seasons, while summer is a little bit higher than fall for office hours (first two columns in Fig. 11). Summer and fall heating energy are almost even for both after hours and office hours (middle two columns in Fig. 11). Afterhours and business hours have also close cooling energy consumption (the 5th and 6th columns in Fig. 11).

**Observation**: the comparison in different granularities shows that there is **no *direct and visible connection between the energy consumption and the occupancy rate***. In other word, a lot of energy is wasted regardless of the actual usage.

To summarize our findings with the energy efficiency modeling and evaluation results in the building-side smart grids and the HVAC system energy consumption, we find that both green buildings like our testbed and most of the conventional buildings are with centralized control and fixed running patterns, which leads to poor energy efficiency in operations though some of them may originally be designed to be "green". We further argue that to enable more energy efficient consumer-side building grids, we need to find a series of networking designs to enable distributed and dynamic control.

We will discuss our major idea and corresponding prototyping and experimentation in the following two sections.

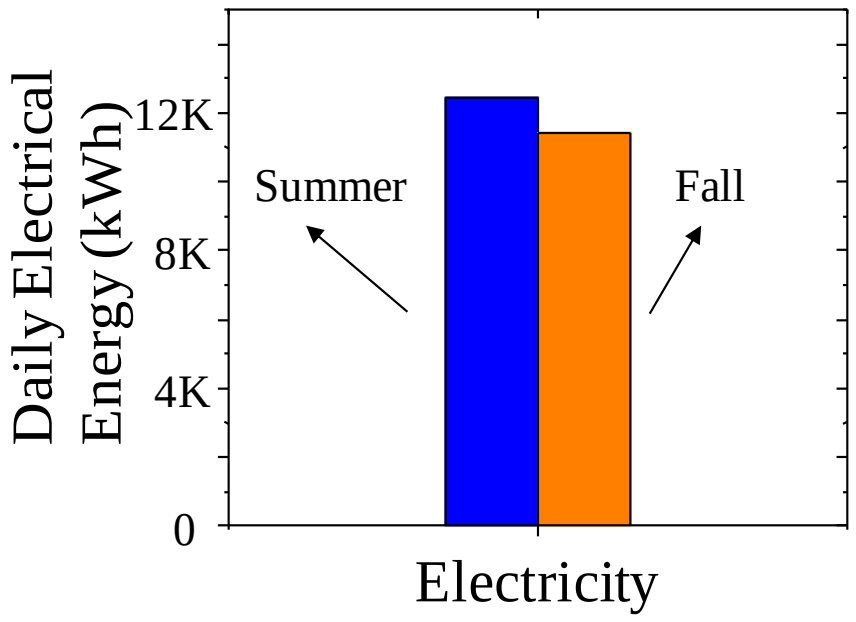

Fig. 9. Summer and Fall daily **electrical** energy consumption comparison

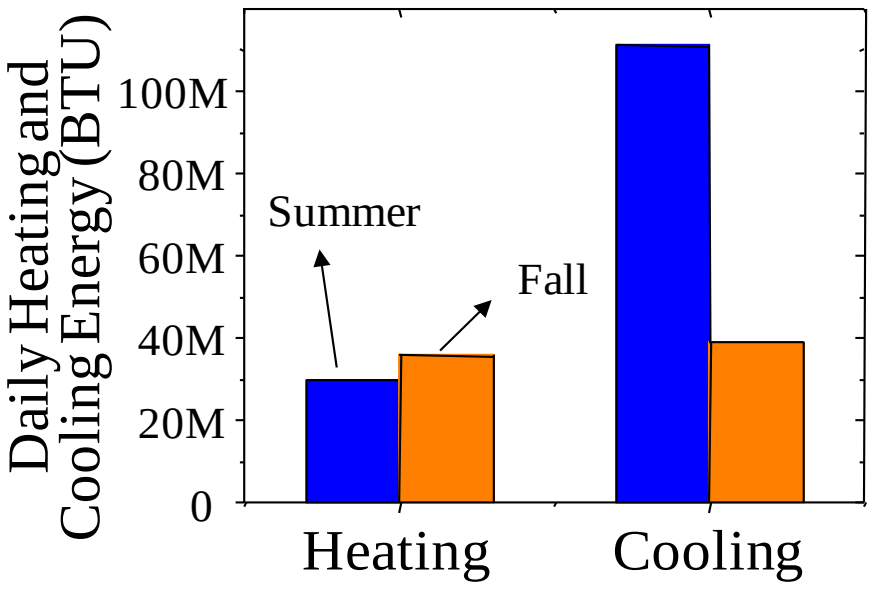

Fig. 10. Summer and Fall daily **heating and cooling** energy consumption comparison

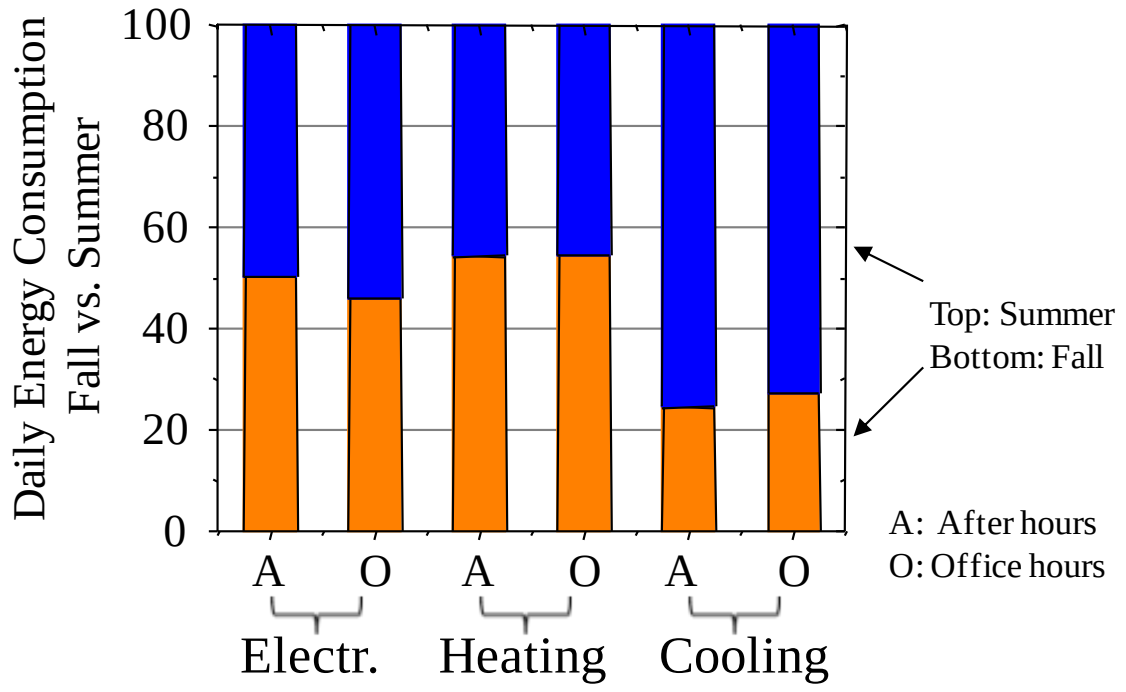

Fig. 11. The daily energy consumption comparison between Summer and Fall, considering after hours and office hours

## III. Smart Location-Based Automated Energy Control Framework

In this section, we present our smart location-based automated energy control IoT framework. This section is mostly based on our contributions in [2].

### A. *Overall Structure*

There are multiple design components and aspects which interact with each other and form a complete framework of our idea to fulfill the goals. We envision an occupant oriented and involved networked system and depict it in Fig. 12.

The key design components include: mobile devices based distributed energy monitoring and remote control, location application on smartphone, multi-source energy-saving policies and strategies, cloud computing platform based data storage and application, and energy data modeling and strategy formation. We discuss these below.

### B. *Smart Mobile IoT Devices as Remote Controls*

In the last several years, smart mobile devices have become very popular. Smartphones generally have multiple networking interfaces such as 3G, WiFi, WiMAX, Bluetooth, and have multiple sensors including GPS sensors. Because of various connectivity provisions and global accessibility to the Internet, they are suitable for use in any system that needs humans' online participation or interaction. The "Internet of Things" [7] trend makes the cost even lower and the sensors are connected to the Internet at all time.

Smartphones are ideal for monitoring, controlling, and managing the energy control systems remotely from anywhere at any time. After appropriate authentication and authorization, the occupants are allowed to modify and change their energy-saving policies online by interacting with the policy servers of their office and residential buildings. Such design allows dynamic changes to the energy-saving policies and offers better flexibility to the occupants. It can be a good complement to the general policy decision process based on the modeling results. Such an "app" can be easily developed for the smartphone based on the web technology.

### C. *Multi-source Energy-saving Policies Hierarchy*

In a real environment, various parts of an organization, such as campus, building, department, and labs may be in charge of different components of a building. Each of these may have their own policies and requirements that need to be taken care of in controlling the energy consumption. Even in a single home building, locations of multiple family members and their preferences need to be taken into account. Therefore, in our location based automatic control scheme, we add policies coming from these levels of control hierarchy.

Fig. 13 shows an example of the policy hierarchy. As shown, there may be a tree-like structure for the building control plane in which there are policy servers enforcing the energy-saving policies covering different levels. This also applies to the residential buildings in which the tree structure may be relatively simple. The mobile users can be connected to the Internet through smartphone, tablet, or even laptop with WiFi connections. In the example shown in Fig. 13, the mobile smartphone holder leaves the home building and travels towards his office building. The movement and location changes will trigger the policy servers to adjust the energy-saving policies for both buildings accordingly. The action steps are denoted as "①②③" in the figure.

In our previous research on next generation Internet [8, 9] as well as the policy-oriented Internet architecture [10], we have experimented with several policy based control schemes. We apply similar ideas to the building and community environments. In particular, each control region can be defined as a "realm" [11] which is managed by a realm manager (also a policy server in our building testbed). Energy control policies may span multiple realms and sometime conflicts may have to be resolved.

### D. *Mobile Device Location-Based Automatic Control*

Almost all phones can determine their location by referring to signal strengths from various transmission towers. New generations of smartphones can provide localization much more precisely with embedded GPS chips. We use this location information in designing automatic control policies that can turn on/turn off energy consuming devices at home or office depending upon the location and direction of movement of the user. By doing so, a dynamic and flexible policy can be applied which satisfies the user's preferences for energy saving and comfort. An "App" on the device can automatically enforce these desired policies.

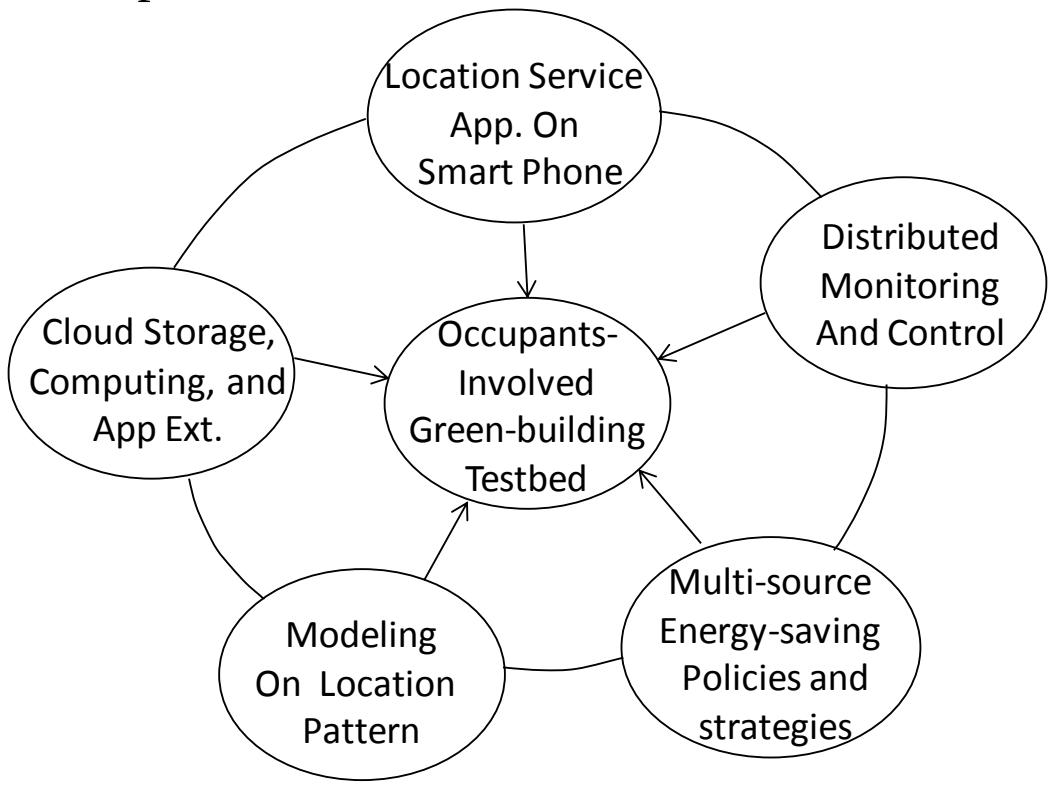

Fig. 12. Overall structure of our design with components and their interaction

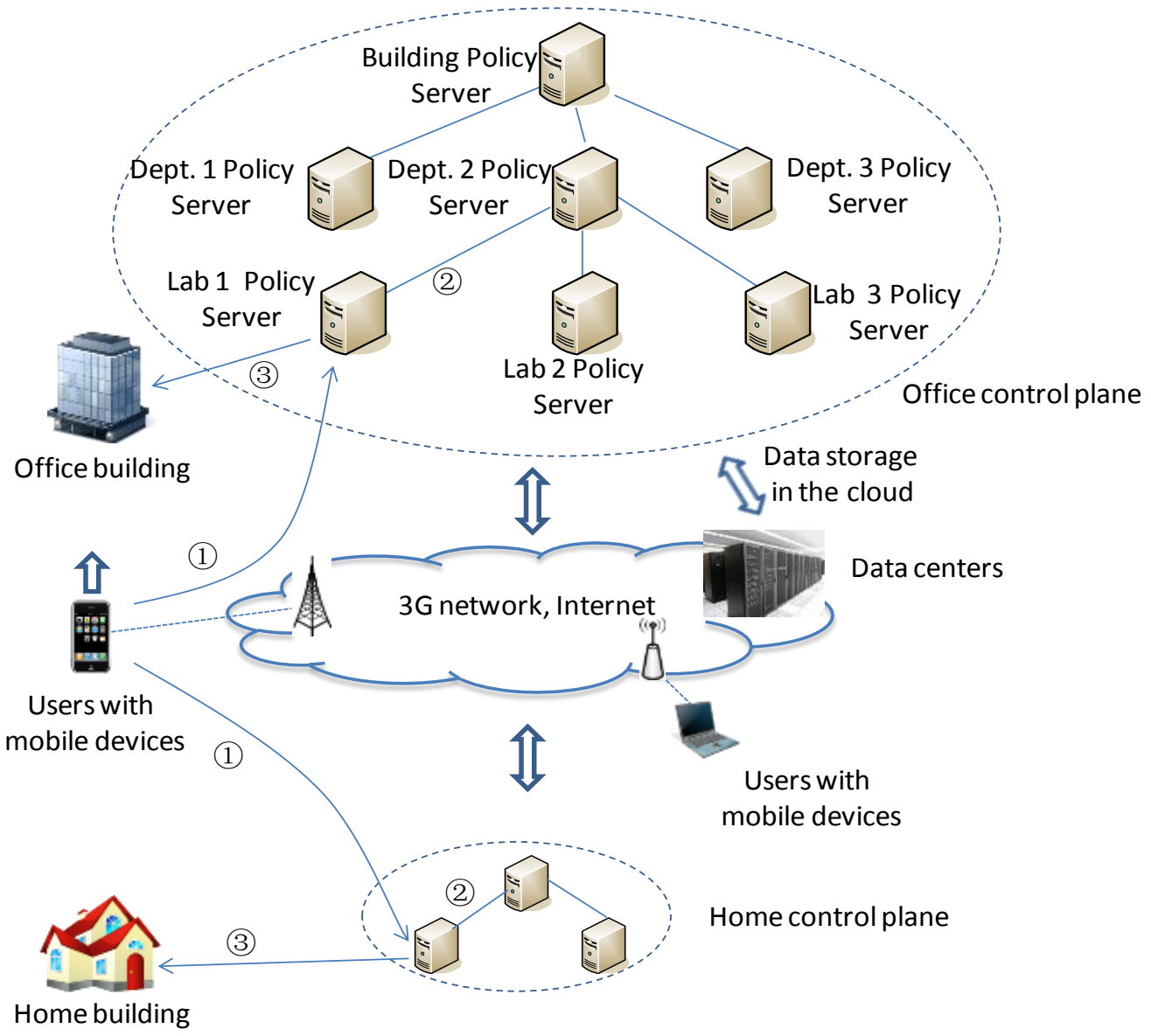

Fig. 13. Example dynamic multi-source energy-saving policy adjustment by the mobile devices

With the help of the location-aware mobile devices, these dynamic adjustment policies could also enable the cooperation and interaction among different buildings. For example, when the location detection daemon on the user's smartphone detects that the user has moved out of a threshold distance range from his home building and is moving into a threshold distance range of his office building, then a message is sent to a centralized server to trigger the policy control process. The office building room owned by the user will start pre-heating/cooling to prepare a user-customized or optimized working environment, while the message also triggers the home building to transit into an energy-saving mode.

### E. *Cloud Computing and Storage*

Cloud computing has become very promising in the last few years. We have two basic kinds of jobs which need the cloud-computing platform: (1) The cloud-based data storage, and (2) the cloud-based modeling and analysis computation. We have a preliminary design of how to integrate the system into the cloud computing platform. As shown in Fig. 14, the cloud provides the basic data storage and retrieval service for the logged building energy consumption data. Computation-intense modeling and analysis jobs are mostly done in the cloud. The communication layer provides configurability, reliability, and security for the network communication between the cloud and the client. The middle layer in Fig. 14 is for cloud application development by using the open API provided by the cloud providers such as Google App Engine. The reason we incorporate this layer in our design is that it can alleviate the overhead to develop the cloud application and accelerate our application development and deployment process. It also becomes much easier to integrate other services using the same platform (such as authentication services, email services and user interfaces) to the application on demand and make the development of a cloud application a less complicated task. The top layer is the application layer. We are researching and developing a user-friendly prototype web-based user interface and application for the building environment, which can be easily configured and managed by the remote client.

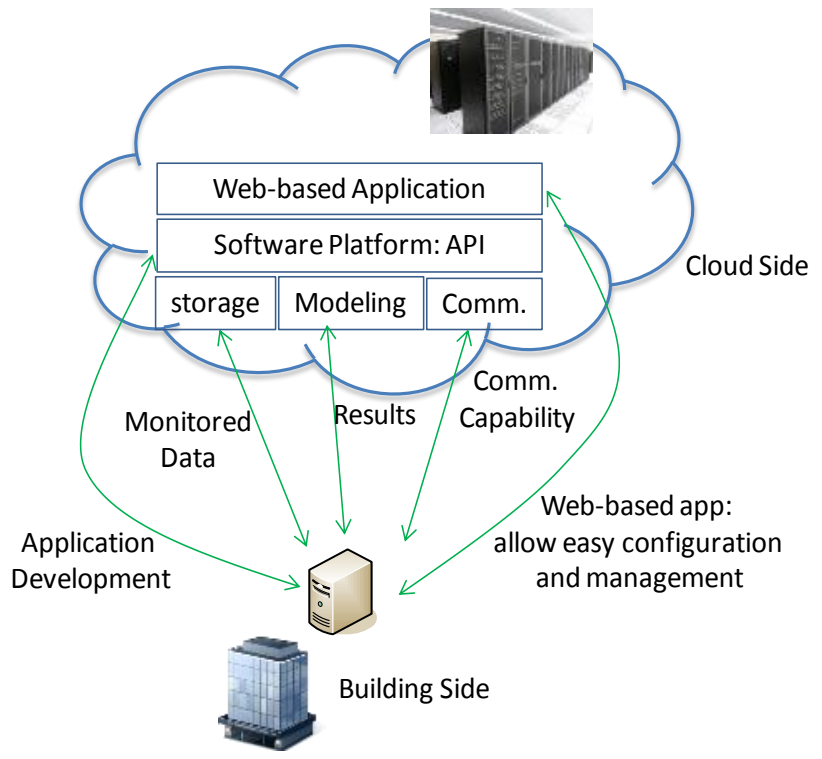

Fig. 14. Cloud computing components and interaction with the building side servers

## IV. Prototype System, Experimental Results, and Multi-scale Energy Proportionality

This section focuses on the prototype system we build to prove the effectiveness of the idea proposed in Section III.

### A. *IoT Prototype Description*

In this IoT prototype system, we implemented a simple scenario involving a user associated to two groups of electrical appliances: those in his/her home apartment and those in his/her office room. It is a simplified scenario of what is shown in Fig. 13. Our goal is to provide users the ability to dynamically adjust and control their devices across two buildings. The basic function is to enable the server to detect the user's location changes and trigger the energy policy changes by turning on/off the electrical appliances in both buildings associated to the user. By doing this, we essentially enable users to control and implement their own energy policies in real time, and enable their energy consumption to be proportional to their actual usage.

Note that in this simple prototype, we only implement the case involving only one user with control devices in two buildings (the user's office building and home building). In other words, we implement this small-scale proof-of-concept system and compare the energy saving with the case that is without the new design. After proving the effectiveness, then we could generalize it into a larger scale. In the future work, we plan to test the case with multiple users controlling their devices simultaneously by which we could show results with a larger scale energy saving.

#### 1) *Hardware and networking structure*

In the prototype system, the hardware systems that we use include the "Kill-A-Watt™" electrical meters [14], WeMo™ control devices [15], servers in each building which act as both web daemon server and in-building controller, WiFi routers, and smart devices with location sensors (GlobalSat GPS module).

The networking structure of the prototype system of the home building side is shown in Fig. 15. The basic function is that a smart mobile device with a location sensor keeps sending its location data back to the web servers inside the home building and the office building. The web daemon servers behind the firewall and NAT (Network Address Translation) are accessed from outside by port mapping technology. It also calculates the distance between it and the mobile devices to decide if the distance passes a specific threshold to trigger energy policy changes in either of the buildings. If it does, then it initiates the controller to send instructions to turn on/off specific devices in its territory according to the energy policies.

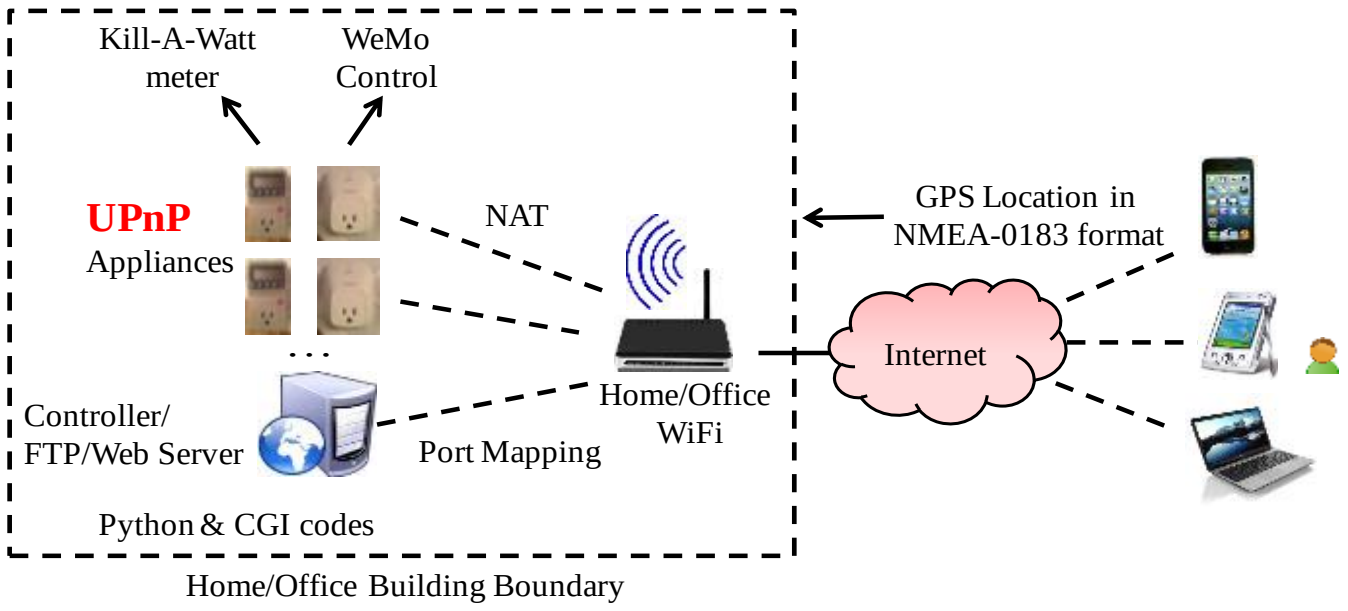

Fig. 15. Prototype system networking structure

#### 2) *Software*

The software part includes the software for GPS location data recording and sending to the web server in NMEA (National Marine Electronics Association) 0183 compliant format, and the WiFi router's configuration and management software which provides a port mapping service for web access from outside of the NAT. The web server is programmed with CGI (Common Gateway Interface) scripts to execute Python codes controlling the WeMo devices through UPnP (Universal Plug and Play) protocol. Besides the location based automated control, these software parts working together with the hardware also enable the devices in both buildings to be controllable from Internet in real time through smart devices.

### B. *Experiments and Results*

We first measure the baseline electricity appliances' power associated with the user in both buildings. The major appliances in the home building of the prototype system and their baseline power measurement and estimation are shown in Table II. Note that in this prototype system, we primarily focus on electricity appliances, though in the real case, HVAC can be a significant energy consuming source worth applying dynamic control to make a difference in improving energy efficiency. Similarly, the appliances in office room and its baseline power measurements are shown in Table III.

To compare and quantify the real savings of our prototype system, we divide the users' energy usage into three potential modes: ***luxury mode, moderate mode***, and ***frugal mode***. For each mode, we estimate how much energy will be consumed on a daily basis. The estimation results for home and office are shown in Table IV and Table V respectively, which also explain the three modes.

Table II. Home electricity appliances' baseline power measurements

| Type | Lighting | | Refrigerator (GE) | Microwave Stove(Philips) | Laptop (Mac Pro 15'') | HVAC |
|---|---|---|---|---|---|---|
| Items | Porch: | 54W | Start: 200W, gradually to 170W<br>Compressor work for 9 min, stop for 9 min | 1.3kW | Normal: 41W<br>Active or charging: 60W | N/A |
| | Bedroom: | 18*2 = 36W | | | | |
| | Living Room: | 54*2+42 = 150W | | | | |
| | Kitchen: | 52*5 = 260W | | | | |
| | Bathroom: | 54W | | | | |
| Avg. Power | | 550W | 185/2 W | 1.3kW | 50W | |

Table III. Office room electricity appliances' baseline power measurements

| Type | Lighting | Desktop | Laptop (Mac Pro 15'') | HVAC |
|---|---|---|---|---|
| Items | 32W * 6 = 192W | Host:<br>Boot – 110W<br>Normal – 67W | Normal: 41W<br>Active or charging: 60W | N/A |
| | | Monitor:<br>Normal—72W, Active—80~90W | | |
| Avg. Power | 192W | 160W | 50W | |

Table IV. Daily home electricity consumption estimation of three modes

| | Lighting | Refrig. | Microwave | Laptop | HVAC |
|---|---|---|---|---|---|
| Luxury Mode (user is energy insensitive) | Always ON except sleeping | Constantly, 185w/2*24 = 2.22 kWh | Constantly, 1.3kw*0.05= 0.065 kWh | Always ON at home | N/A |
| | 550w*24*2/3= 8.8kWh | | | 50w*24*2/3= 0.8kWh | |
| Moderate Mode | Only ON when at home awake | | | Only ON when at home awake | |
| | 550w*24*1/3= 4.4 kWh | | | 50w*24*1/3= 0.4 kWh | |
| Frugal Mode (user is energy sensitive) | Only 60% ON when at home awake | | | Only 60% ON when at home awake | |
| | 4.4 *0.6 =2.64kWh | | | 0.4*0.6 = 0.24kWh | |
| Total | Luxury: 11.90kWh ; | Moderate: 7.09kWh; | | Frugal: 5.17kWh | |

*Assuming 8 hours working in office, 8 hours at home awake, and 8 hours sleeping

Then we apply our location based solution and dynamically control the appliances in both home and office to reduce the energy waste and maximize the energy efficiency. We track and record the location of the user in 24 hours' period and apply dynamic control and policy changes in both home and office. The location history shown in Google map is in Fig. 16.

Note that for the detailed turning on/off policy changes, we consider some real-life limitations. For example, in our testbed, we did not control the on/off status of the refrigerator. We only apply changes to those devices such as lighting bulbs, desktop, and laptop, whose on/off status do not directly affect the normal living of the human being.

Table V. Daily office electricity consumption estimation of three modes

| | Lighting | Desktop | Laptop | HVAC |
|---|---|---|---|---|
| Luxury Mode (user is energy insensitive) | Always ON when at office<br>192W*8= 1.54kWh | Always ON 24/7 | Always ON when at office | N/A |
| | | 160W*24 =3.84kWh | 50W*24*1/3= 0.4kWh | |
| Moderate Mode | | Only ON when at office | Only 50% ON when at office | |
| | | 160W*24*1/3 =1.28kWh | 0.4*0.5= 0.2 kWh | |
| Frugal Mode (user is energy sensitive) | | Only 60% ON when at office | OFF when at office, use desktop | |
| | | 1.28*0.6=0.77kWh | 0kWh | |
| Total | Luxury: 5.78kWh ; | Moderate: 3.02kWh; | Frugal: 2.31kWh | |

*Assuming 8 hours working in office, 8 hours at home awake, and 8 hours sleeping

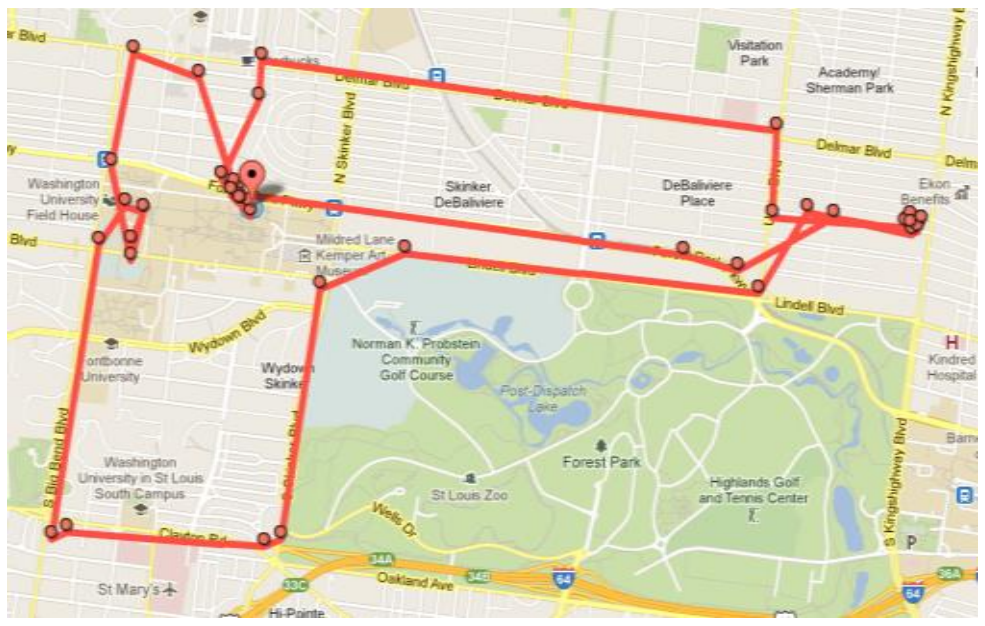

Fig. 16. Location history in a 24 hours' period

In the real activity trace of our experiments, it shows that approximately the user spent 14 hours at home in which 8 hours for sleep, 2 hours for lunch and rest, and 4 hours for working at home. The total recorded real energy consumption at home during this period is 5.285 kWh, which includes 2.7kWh for lighting, 2.22kWh for refrigerator, 0.065kWh for microwave stove, 0.3kWh for laptop.

Also, during this period, the appliances in office room are kept in "OFF" status by the control server of the prototype system. Location history also shows that about 6 hours are spent in office and almost half of the time the desktop is used and for the other half time the laptop is used. The real total energy consumption at office is 2.26kWh, which includes 1.15kWh for lighting, 0.96kWh for desktop, and 0.15kWh for laptop. For the remaining 4 hours, the user is not at home/office, and all the devices are in "OFF" status, except the refrigerator at home.

Thus, for comparison, we put the real recorded energy consumption data after applying our ideas together with the energy consumption estimation results of the three modes, to demonstrate how much energy can be saved. The results are shown in Fig. 17. The simple takeaway message is that the real energy consumption of the prototype system after applying our location based idea is ***very close to the frugal mode's energy consumption***. It means that with our new idea, general users will enjoy luxury living style without special care or changes and they will pay what frugal mode users pay.

### *C. System Implementation and Integration Challenges*

During the system prototyping process, we met multiple challenges and found corresponding solutions to address these issues. We discuss them here as follows.

*1) Extending Control Device Functions*

The first challenge for our system implementation was that we needed to find appropriate devices and methods to carry out the control functions. More importantly, these methods should not be limited by a specific software or platform. It is desirable to allow the devices to be controllable from anywhere at any time through the general PC platforms while not limited by those dedicated platforms such as iOS or Android. Therefore, we "hacked" the WeMo devices by wrapping up and extending some standard uPnP protocol control module and successfully realized the above functions.

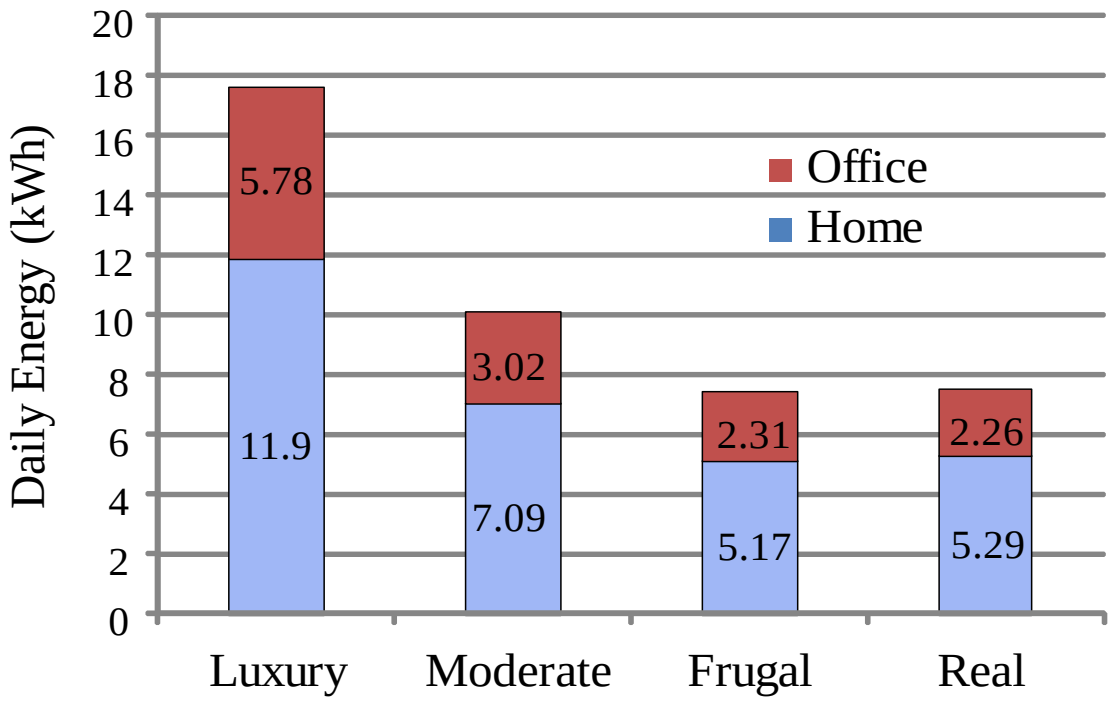

Fig. 17. Comparison of the real energy consumption after applying our idea with the three modes' energy estimation

*2) Integrating Multiple Components*

As shown in Fig. 15, in our prototype, we have a series of sub-components to be integrated into the prototype system. For example, the client mobile devices with GPS sensor, the client data uploading module, the data and web server, the controller, the appliances under control, and the WiFi router with port mapping function. It has been a challenge to organize and integrate these heterogeneous devices to work together as a coherent system. In our prototype, we successfully addressed this challenge and implemented a working system.

*3) GPS Data Fast Parsing*

In our prototype system, we used a GlobalSat GPS sensor, and the coordinate data were encoded in NMEA-0183 data format. The GPS sensor generates a large amount of coordinate data in real time which requires the server to receive them and perform fast parsing to get the accurate coordinates of the mobile user. We implemented the function in Python to fulfill this task, which also includes function implementing the distance calculation and the threshold comparison before triggering the energy policy changes in both sides of the buildings in the testbed.

*4) Handling GPS Location Inaccuracy*

Another challenge was that the GPS coordinate data generated by the GlobalSat sensor has its own accuracy limit, which means that even the mobile user does not move, the generated coordinate data may vary. The following Fig. 18 approximately demonstrates this effect. In the figure, we can see that even for an unmoved mobile user, its coordinate data trace may vary by 20 meters given the inaccuracy of the GPS sensors.

Therefore, to overcome such difficulty, we tuned the threshold distance and did comprehensive experiments before reaching a threshold that could minimize the false alarm or false positive rates. Our results proved to be satisfactory.

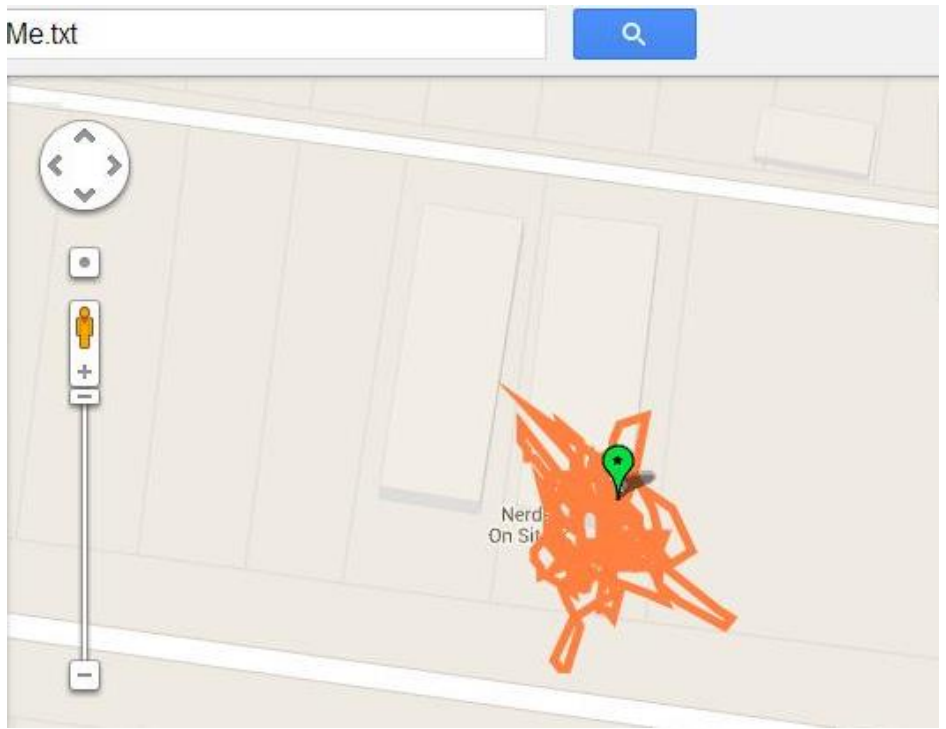

Fig. 18. GPS location trace with an unmoved mobile user

## D. Multi-scale Energy Proportionality

To summarize the experiments and results, we can see that the effect of energy saving is conspicuous, although given different types of buildings and occupants' energy using habits, there can be different degrees of savings.

The above prototype system experiments vividly illustrate how ***user-scale energy proportionality*** is realized by using networking and computing technologies, since the user's energy consumption becomes approximately proportional to his/her actual usage. Besides user-level energy proportionality, applying similar idea into multi-scale organizations in a building, as described in Section III, it virtually enables both ***organization-level energy proportionality and building-level energy proportionality***. Specifically, when a particular user controls and adjusts the appliance policies under his/her territory, he/she has to follow specialized group policies enforced by the organization such as a laboratory or a department. The organization also enforces the policies for publicly shared parts such as HVAC, lighting, fire and safety, and elevator systems. It could designate special working staff to control and apply energy proportionality for these subsystems. The laboratory or department ***aggregates*** each user' energy proportionality and the publicly shared subsystems' energy proportionality, to achieve an organization scale energy proportionality. Similarly, the idea can be generalized to building level since it basically aggregates multiple organizations inside the building and multiple public subsystems working for all the organizations in the building.

Achieving multi-scale energy proportionality has profound economic impacts to the society in terms of avoiding huge energy waste and saving costs for users and organizations. The benefits would make a huge difference if the idea gets broadly implemented and deployed. The networking and computing technologies used for the system enable the buildings running and operation to be more intelligent and efficient, and in an automated manner without manual intervention, which is also very important goal for future smart home/office applications. All the stakeholders including the common occupants and users, organizations' authorities, and buildings' owners and tenants could have total flexible control over their own energy policies, which is a very promising feature for our proposed idea. Moreover, the proposed system also involves every user and organization to participate in the energy saving efforts, which is potentially a very good training and education method to encourage everyone to study and participate in resolving global climate and sustainability issues in everyday life. Further incorporation of social network plugin into the smart mobile phone based energy control platform would generate even broader impacts [2].

## V. Related Work

Due to the multi-disciplinary essence of the research topic, the related work covers a range of different areas. We discuss a few of them briefly. Limited by the space, a longer list of the related work can be found in papers [2, 3].

First related area is the building energy simulation. Many building simulation software take building parameters as input and after processing, output estimated energy usage [16]. A popular one is "EnergyPlus" [12] provided by the Department of Energy (DOE). In comparison, monitoring building energy consumption with real network system and analyzing the real energy consumption data can be more effective.

Second area is the climate effect models research in which a lot of existing work is about the relationship between energy consumption and climate or weather factors [17, 18]. The related research consists of: (1) simulating the heat transfer processes and building structures (envelope, tree shelter, etc.) to find how the climate can impact building energy efficiency; (2) study of solar effects on heat and mass transfer and their impacts. Moreover, most of them are currently based on theoretical thermal calculations and simulations and very few of them are using the actual building energy data and research on how to reduce the energy consumption by incorporating occupants' participation. A longer reference list of related efforts can be found at [19].

Third area is the application of Wireless Sensor Network (WSN) technologies into the building environment and experiments on a specific subsystem like lighting and thermostat. For example, it is used to sense and control the lights according to the detection results of the sunlight for a building based on human activities, to monitor the electrical energy consumption, to log the human activities and to adjust the HVAC working time to provide better comfort, etc. There are already a lot of such experimental researches. Typical ones are in [20, 21] and more can be found from the list at [19]. However, WSN were originally designed for other purposes but they are able to provide a good complement to other sensing and metering technologies in the building environment.

Last related category lies in the information and computer science technologies. The iPhone, Android, and Windows Phones provide similar open platforms to develop versatile smart applications with multiple sensors including GPS. So far, there are not many wide-scale applications on building environment energy auditing and control. Cloud computing [13] certainly is a very hot topic related to our research.

## VI. Conclusions

In this paper, we added new contributions besides summarizing our previous work regarding the IoT framework for smart energy in buildings. The work includes: (1) energy consumption data analysis of the green building testbed, (2) new smart location-based automated energy control framework designs, and (3) experimental prototype that applies IoT networking and computing technologies to improve the energy efficiency in buildings. We put them into a complete three-step research and added significant new contributions proving the ideas and concepts we proposed. By building this IoT framework in smart homes or offices, we aim to enable not only multi-scale energy proportionality, but also create an intelligent home space which is an important part of the future smart world. We envision that the idea will provide not only significant economic benefits but also huge social benefits in terms of global sustainability.

**Jianli Pan** [M] is currently an assistant professor of Computer Science in the Department of Mathematics and Computer Science at the University of Missouri - St. Louis, MO USA. Previously he obtained his Ph.D. degree in Computer Engineering from Washington University in Saint Louis, MO USA. He also holds two master degrees from Washington University in Saint Louis and Beijing University of Posts and Telecommunications, respectively. His current research is on the Future Internet Architecture, mobile and cloud computing, network virtualization, Internet of Things, cyber security, and smart energy in buildings and smart grids.

**Raj Jain** [F] is a Fellow of ACM and AAAS, a winner of the ACM SIGCOMM Test of Time Award and CDAC-ACCS Foundation Award 2009, and ranks among the top 100 in CiteseerX's list of Most Cited Authors in Computer Science. He is currently a professor in the Department of Computer Science and Engineering at Washington University. Previously, he was one of the cofounders of Nayna Networks, Inc., a next-generation telecommunications systems company in San Jose, California. He was a senior consulting engineer at Digital Equipment Corporation in Littleton, Massachusetts, and then a professor of computer and information sciences at Ohio State University, Columbus. He is the author of *Art of Computer Systems Performance Analysis*, which won the 1991 Best-Advanced How-to Book, Systems Award from the Computer Press Association.

**Subharthi Paul** [M] is currently a member of technical staff in Cisco, INC. He received his Ph.D. from Washington University in Saint Louis, MO USA. Before that, he received his B.S. degree from the University of Delhi, India, and his Master's degree in software engineering from Jadavpur University, Kolkata, India. His primary research interests are in the area of future Internet architectures.

**Tam Vu** [M] is currently an assistant professor in the Department of Computer Science and Engineering at the University of Colorado, Denver. He holds a Ph.D. from the Department of Computer Science at Rutgers University. He got the Best Paper Awards at the Annual International Conference on Mobile Computing and Networking (MobiCom) in 2011 and 2012. His current research is on mobile systems and wireless networks and their various properties.

**Abusayeed Saifullah** [M] is an assistant professor in the Computer Science Department at Missouri University of Science & Technology. He received PhD in 2014 from the Computer Science and Engineering Department at Washington University in St Louis. His research primarily concerns cyber-physical systems with contributions spanning real-time systems, embedded systems, wireless sensor networks, and parallel and distributed computing. He received the Best Paper Award at RTSS '14, the Best Student Paper Award at RTSS '11, Best Paper Nomination at RTAS' 12, and the Best Student Paper Award at ISPA '07.

**Mo Sha** [M] is currently an assistant professor in the division of science of the Governors State University, Illinois. He obtained his Ph.D. degree from Washington University in Saint Louis.